\documentclass[aps, pra, onecolumn, a4paper, 12pt, amsmath, preprintnumbers, nofootinbib, tightenlines, longbibliography, notitlepage]{revtex4-1}\pdfoutput=1
\usepackage[english]{babel}
\usepackage{bbm}
\usepackage{graphicx}
\usepackage{hyperref}

\graphicspath{{pics/}}

\newcommand{\floor}[1]{\left\lfloor#1\right\rfloor}
\newcommand{\hh}[1]{\left(#1\right)}
\newcommand{\bb}[1]{\big[#1\big]}
\newcommand{\bhh}[1]{\big(#1\big)}

\newcommand{\RR}{\mathbbm{R}}
\newcommand{\CC}{\mathbbm{C}}

\newcommand{\ZZ}{\mathbbm{Z}}
\newcommand{\ii}{\mathbbm{i}}
\newcommand{\ee}{\mathbbm{e}}

\newcommand{\Id}{\mathbbm{I}}

\newcommand{\PSL}[1]{\textrm{PSL}(#1)}

\newcommand{\md}{\mathrm{d}}

\hypersetup{pdfauthor={Maarten van de Meent},pdftitle={Gravitational Waves from Piecewise Flat Gravity},pdfkeywords={ Locally flat - Straight strings - Classical general relativity  - Gravitational plane waves}}

\begin{document}
%Short title:Exact Piecewise Flat Gravitational Waves
\preprint{ITP-UU-11/22}
\preprint{SPIN-11/17}
\title{Exact Piecewise Flat Gravitational Waves}

\author{Maarten \surname{van de Meent}}
\affiliation{Institute for Theoretical Physics and Spinoza Institute,\\ Utrecht University,\\ P.O. Box 80.195, 3508 TD Utrecht, the Netherlands}
\email{M.vandeMeent@uu.nl}

\date{\today}

\begin{abstract}
We generalize our previous linear result \cite{meent:2010b} in obtaining gravitational waves from our piecewise flat model for gravity in 3+1 dimensions to exact piecewise flat configurations describing exact planar gravitational waves. We show explicitly how to construct a piecewise flat spacetime that describes an impulsive plane wavefront. From these wavefronts more general plane waves may be constructed. Further remarks are made on how this construction may be extended to non-plane (e.g. cylindrical) waves.
\end{abstract}
\maketitle

\section{Introduction}
In previous articles \cite{hooft:2008,meent:2010} 't Hooft and the author have introduced a piecewise flat model for gravity in 3+1 dimensions, motivated by general relativity in 2+1 dimensions. In 2+1 dimensions, the Riemann tensor is equivalent to the Einstein tensor. Consequently, the Einstein equation completely and locally determines the geometry in terms of the matter content. In particular, empty space is flat and has no local purely gravitational degrees of freedom, facilitating quantization of pure gravity \cite{Witten:1988hc,Deser:1989}. A system of point particles generates a piecewise flat geometry, with the particles represented by moving conical curvature defects \cite{Deser:1983tn}. Such a configuration has locally only a finite number of degrees of freedom and may be quantized \cite{Waelbroeck:1994iy, hooft:1996uc}.

One obstruction to quantizing gravity in 3+1 dimensions is the presence of local gravitational degrees of freedom in the form of a spin-2 graviton field, which is non-renormalizable when quantized perturbatively. The idea behind the model discussed in  \cite{hooft:2008,meent:2010} is to take a page from gravity in 2+1 dimensions, and eliminate all local degrees of freedom by enforcing that empty space should be completely flat. Local degrees of freedom are introduced as conical curvature defects. If we insist that the Einstein equation be satisfied, then the flatness of empty space implies that these defects are flat co-dimension 2 surfaces. Since these defects are interpreted as physical degrees of freedom, causality requires that only timelike surfaces appear. This distinguishes the model from other piecewise flat approaches to gravity such as Regge calculus \cite{regge:1961,RW:2000}.

The additional assumption that empty space is flat appears quite limiting. In particular, at first sight it appears that it completely eliminates the possibility of gravitational waves. However, in the model the propagating line defects represent both the matter and gravitational degrees of freedom. For this reason the model allows both defects that contribute positively and negatively to the curvature.\footnote{If the defects are interpreted purely as matter then the negative curvature defects would correspond to matter with negative energy density, and should not appear.} The appearance of positive and negative curvature defects opens the possibility of geometries for which the Einstein curvature vanishes when averaged over larger scales, while the Riemann curvature does not. Some such geometries could  be interpreted as emergent gravitational waves. 

In our previous article \cite{meent:2010b}, we studied this possibility in the linearized limit of our piecewise flat gravity model. The two main results were that: 1) Configurations of moving defects can produce the energy--momentum distribution of an arbitrary dust cloud of point particles, despite the spacial extent of the defect lines; 2) There exists a large family of linear defect configurations with vanishing energy--momentum, whose metric perturbation corresponds with linear gravitational waves in general relativity. At the linear level, the model is able to reproduce gravitational waves as an emergent feature at larger scales.

In this article we extend this result to exact gravitational waves. In contrast to linearized general relativity, it is not possible to erive completely  general exact vacuum solutions of general relativity. Deriving exact solutions requires some form of additional symmetry.  For example, it is known how to construct completely general gravitational waves with cylindrical symmetry \cite{Weber:1957}, and how to construct exact gravitational plane waves \cite{bondi:1959}.

Here we will focus on plane gravitational waves and show how to construct a family of increasingly fine discrete configurations of defects that approach an exact gravitational wave in the limit that the configuration becomes continuous.

Our strategy will be to first  consider gravitational plane waves in the linearized limit, and examine what linearized distributions of defects approximate these waves (section \ref{sec:linplanewaves}). Based on the linearized configuration, we will make an ansatz for an exact piecewise flat planar wavefront, which reduces to the linearized distribution in the linear limit. We show that this ansatz is indeed a bona-fide piecewise flat configuration, and that in the limit where the configuration becomes continuous it produces an impulsive gravitational wave solution of general relativity (section \ref{sec:planewavefronts}). In section \ref{sec:genplanewaves} we combine trains of these wavefronts to form general exact gravitational plane wave solutions. Finally, in section \ref{sec:otherwaves} we comment on the construction of non-planar waves and the additional subtleties that those entail.

\section{Linear plane waves}\label{sec:linplanewaves}
For simplicity we shall consider, without loss of generality, plane waves travelling in the positive $z$-direction. Expressed in lightcone coordinates $(u,v,x,y)$,\footnote{In this article, spacetime metrics have signature $(- + + +)$, lightcone coordinates are defined as $u=\tfrac{z+t}{\sqrt{2}}$ and $v=\tfrac{z-t}{\sqrt{2}}$, and we use units such that $c = \hbar = 8\pi G =1$.} such a wave takes the following form in linearized general relativity,
\begin{equation}\label{eq:gwpert}
h_{\mu\nu} = \begin{pmatrix}
0 & 0 & 0 & 0\\
0 & 0 & 0 & 0\\
0 & 0 & h_{+} & h_{\times}\\
0 & 0 & h_{\times} & -h_{+}\\
\end{pmatrix}\ee^{\ii \omega u},
\end{equation}
where $h_{\mu\nu}$ is the perturbation of the metric on the Minkowski background. In our previous article \cite{meent:2010b} we saw how to produce such a metric perturbation in linearized piecewise flat gravity. The basic building block was a configuration we called a ``laminar plane wave of defects''. A laminar plane wave of defects is a space-filling configuration of parallel conical curvature defects (i.e. line defects that have the same orientation and velocity), whose energy density varies as the amplitude of a plane wave with a wave vector perpendicular to the defects. That is, given a point $x^\mu$ in our spacetime there is exactly one defect (spanned by 4-vectors $d^\mu$ and $u^\mu$) in the laminar plane wave configuration that passes through this point, because all the defects in the laminar plane wave are parallel. The linear energy density of this defect is given  by $\Omega \ee^{\ii k_\mu x^\mu}$ with some fixed wave vector $k_\mu$, which is perpendicular to $d^\mu$ and $u^\mu$.

\begin{figure}[btfp]
\includegraphics[width=120mm]{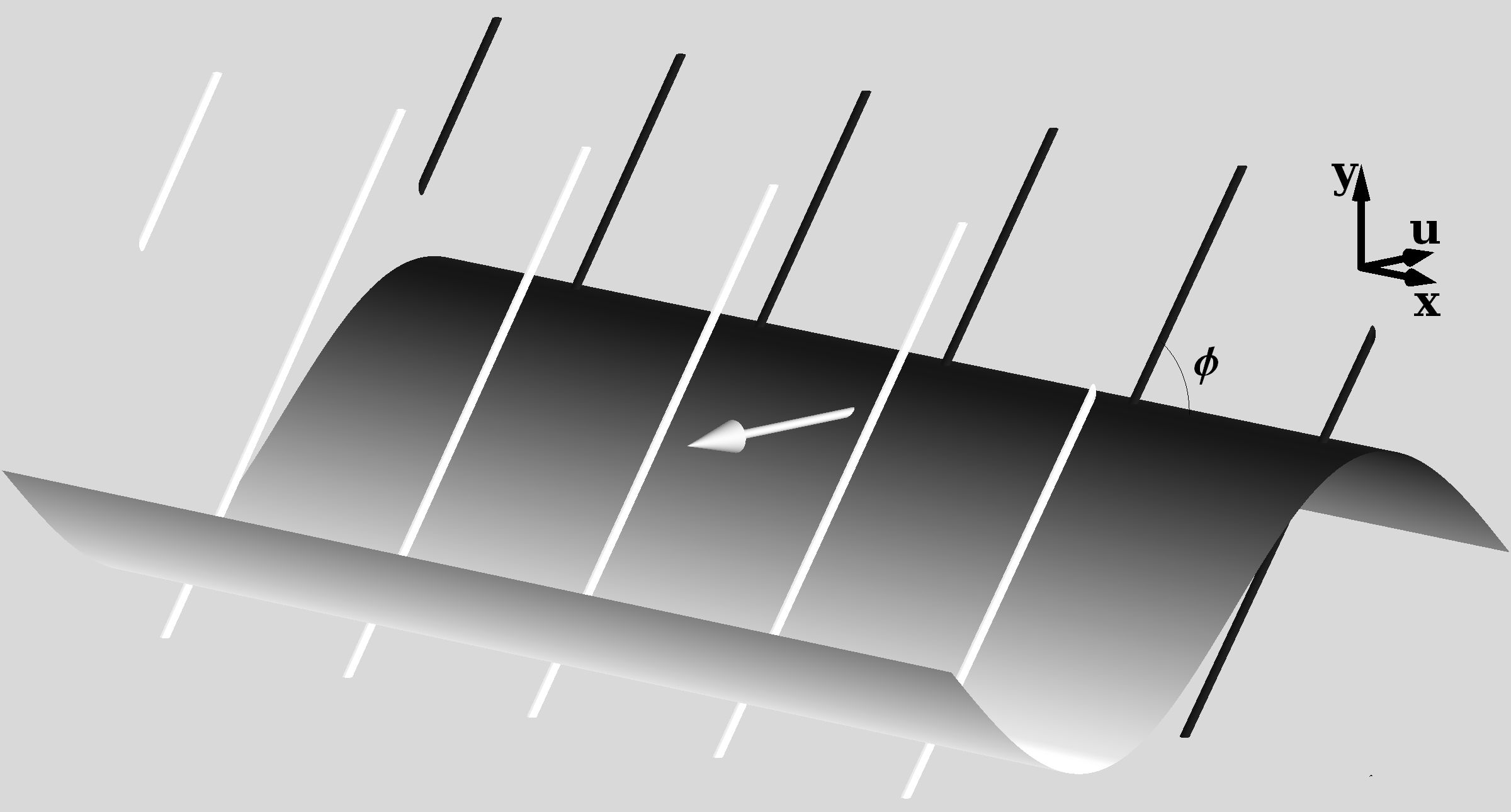}
\caption{A laminar plane wave consists of parallel line defects travelling in the same direction, whose densities vary as a plane wave in the direction of propagation.}\label{fig:laminarplanewave}
\end{figure}

In the linearized limit of our piecewise flat model an arbitrary configuration of defects may be written as a linear combination of laminar plane waves. In  \cite{meent:2010b} we found that a gravitational wave of the form \eqref{eq:gwpert} was a produced as a superposition of laminar plane waves of defects  with the velocity $u^\mu=(1,0,0,1)$ and wave vector $k^\mu=(\omega,0,0,\omega)$. The orientation $d_\mu$ of the defects in these laminar plane waves can then be parametrized by a single angle, $\phi$, which we choose to be the angle between the defect lines and the $x$-axis.

To be a gravitational wave the total energy density amplitude of the superposition of laminar plane waves must vanish. That is, if the energy density amplitude of the laminar plane wave with orientation $\phi$ is called $\Omega(\phi)$, then this condition can be expressed as
\begin{equation}\label{eq:omegacond}
  \int_0^\pi\Omega(\phi) \md\phi = 0.
\end{equation}
It was established in \cite{meent:2010b} that this is the only condition on the superposition of laminar plane waves of defects required for it to describe a gravitational plane wave. The polarization of the resulting gravitational wave was shown to be given by
\begin{equation}
\begin{aligned}
  h_{+} & =  \int_0^\pi\Omega(\phi)\cos(2\phi) \md\phi,\\
 h_{\times} & =  \int_0^\pi\Omega(\phi)\sin(2\phi) \md\phi.
\end{aligned}
\end{equation}
All higher moments of $\Omega(\phi)$ are unconstrained. That is, there is a large family of linear configurations of defects that produce the same linear gravitational wave. We therefore have quite a bit of playing room for finding a linearized configuration that can be interpreted as the limit of some exact configuration of defects.

Geometrically, it would be convenient if $\Omega(\phi)$ has as few discrete components as possible. Condition \eqref{eq:omegacond} implies that we need at least two components. This is, in fact, sufficient to create waves of any polarization. If we take
\begin{equation}
  \Omega(\phi)  =  \alpha\hh{ \delta(\phi-\phi_0) -  \delta(\phi-\phi_0-\tfrac{\pi}{2})},
\end{equation}
then the polarizations are given as $h_{+}= 2\alpha \cos2\phi_0$ and $h_{\times}= 2\alpha \sin2\phi_0$. We thus have perpendicular components of opposite energy, with the amplitude of the defects determining the amplitude of the wave, and the orientation determining the polarization.

\section{Plane wavefronts}\label{sec:planewavefronts}
\begin{figure}[btfp]
\includegraphics[width=120mm]{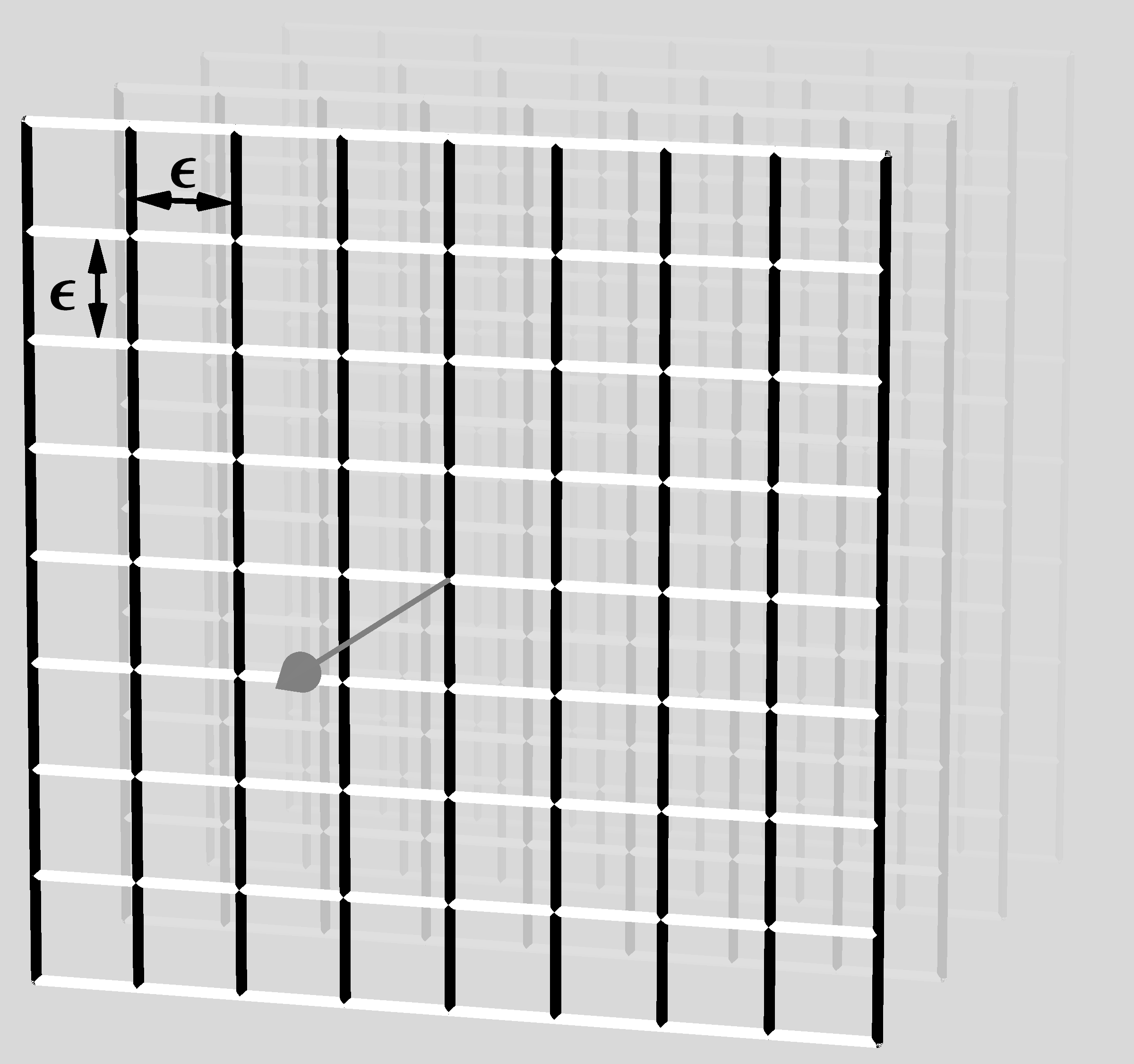}
\caption{A wavefront of perpendicular positive (white) and negative (black) energy line defects.}\label{fig:gridfront}
\end{figure}

This linear configuration prompts us to make an ansatz for an exact configuration of defects. Further discretizing the linear distribution, we guess that a configuration consisting of subsequent grids of perpendicular positive and negative energy massless defects should suffice to approximate a gravitational wave. To analyse this configuration we first consider a wavefront located at $u=\tfrac{z+t}{\sqrt{2}}=0$ consisting of a single grid (see figure \ref{fig:gridfront}).

We first need to check if this ansatz defines a proper configuration of defects. In \cite{hooft:2008} the conditions that should be met at a junction of multiple defects were discussed. These can be phrased in terms of the holonomies of loops around the defect lines. They can be summarized as saying that the product of the holonomies of a set of loops whose concatenation is contractible should be the identity. For the proposed grid configuration the junction conditions must be satisfied at each vertex of the grid.

Since all the horizontal line defects in figure \ref{fig:gridfront} have the same density, the holonomy of each simple loop around a section of a horizontal line defect should be the same, and similarly for the holonomies of simple loops around vertical line defects.   If we call the holonomy of a simple loop around a section of a horizontal grid lines $Q_a$ and the holonomies of the vertical grid lines $Q_b$, then at each vertex the junction condition is

\begin{equation}\label{eq:ewjc}
Q_a Q_b {Q_a}^{-1}{Q_b}^{-1} = \Id.
\end{equation}

The holonomy of a lightlike defect is a null rotation, i.e. a Lorentz transformation that leaves a lightlike and a perpendicular spacelike direction invariant. In the $\PSL{2,\CC}$ representation of the Lorentz group, the holonomy of a general lightlike defect moving in the $z$-direction is represented as
\begin{equation}
Q_z = \begin{pmatrix}
1 & \zeta \\
0 & 1 
\end{pmatrix},
\end{equation}
where $\zeta$ is a complex number. The argument of $\zeta$ gives the direction of the defect, while its modulus gives the energy density of the defect, i.e. the parabolic angle of the null rotation. The junction condition for $n$ lightlike defects moving in the $z$ direction with complex parameters $\zeta_1,\ldots, \zeta_n$ therefore reduces to
\begin{equation}
\zeta_1 +\cdots +\zeta_n =0.
\end{equation}
Furthermore if $Q_a$ has complex parameter $\zeta_a$ then ${Q_a}^{-1}$ has complex parameter $-\zeta_a$. Hence the junction condition for the vertices of the grid \eqref{eq:ewjc} reduces to
\begin{equation}\label{eq:gridjunction}
\zeta_a + \zeta_b - \zeta_a -\zeta_b =0,
\end{equation}
which is true for any $\zeta_a$ and $\zeta_b$. Our ansatz configuration of a grid of perpendicular lightlike defects is therefore a bona-fide configuration of defects for any value of the amplitudes of its components. In particular, it is valid for the particular case we are interested in, where the components have opposite amplitudes.

Our next step is to compare this configuration with an exact plane wave solution of the vacuum Einstein equation. For this we need to find a metric to describe the geometry of the grid of defects.

Away from the grid the spacetime is flat. Hence, if we position the grid at $u=0$, we can choose the metric for $u<0$ to have the familiar Minkowski form,
\begin{equation}\label{eq:lcMinmetric}
\md s^2 = 2\md u \md v+ \md x^2 +\md y^2.
\end{equation}
For simplicity we choose to align the grid with the $x$ and $y$-axes,\footnote{This should correspond to wavefront with ``$+$'' polarization, according to the linear analysis.} such that the positive energy defects are aligned in the $x$-direction and sit at constant values of $y = \epsilon(n  +\tfrac{1}{2})$ for integer values of $n$ and grid spacing $\epsilon$, and the negative energy defects are oriented in the $y$-direction and sit at constant values of $x = \epsilon(m+\tfrac{1}{2})$. 

\begin{figure}[tbfp]
\centering
\includegraphics[width=120mm]{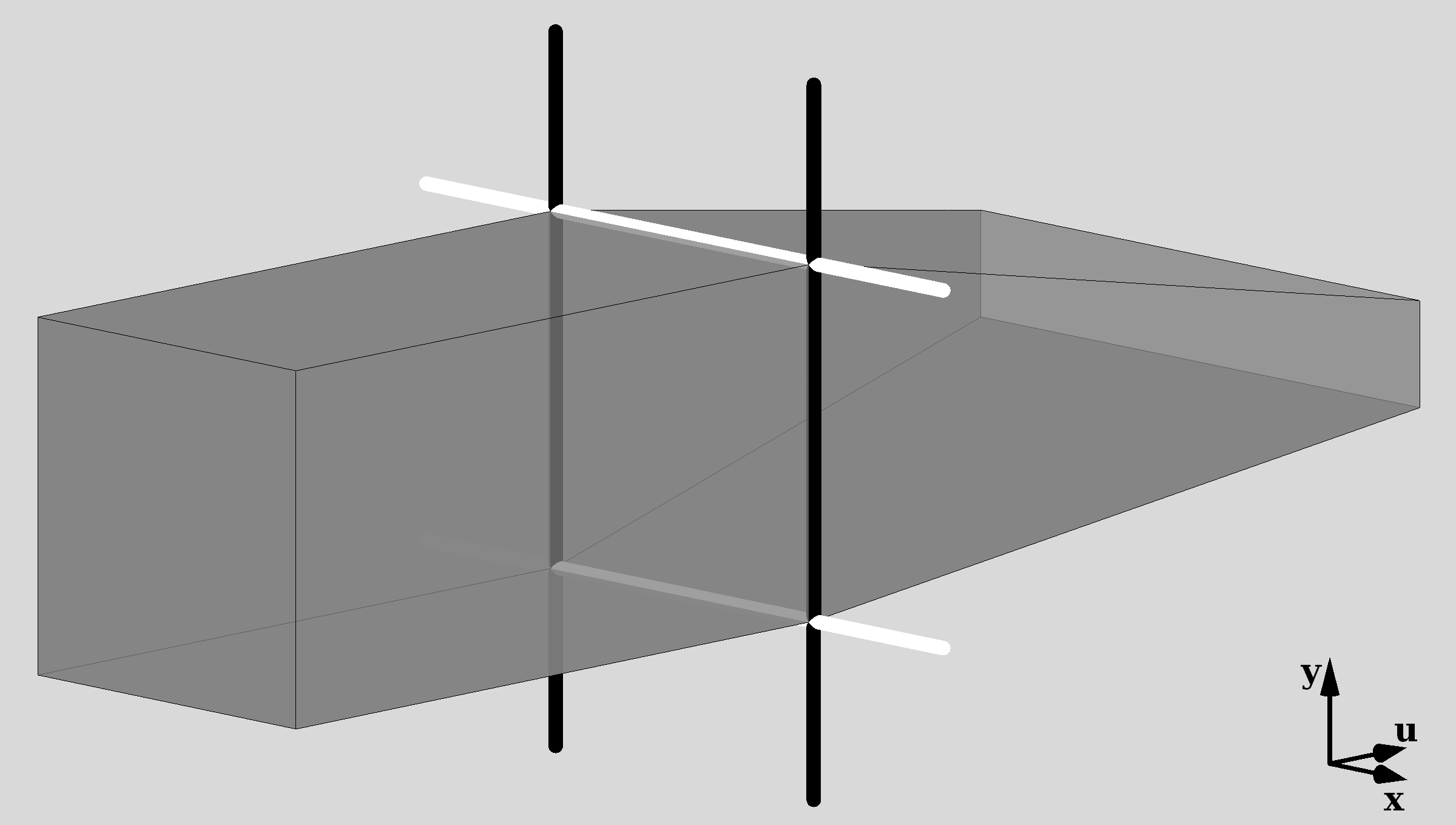}
\caption{The geometry of one ``tube'' of spacetime.The deficit angles of the positive energy defects (white), cause the geometry to be squeezed in the $y$-direction as $u$ increases. Similarly the surplus angles of the negative energy defects (black) cause the geometry to be elongated in the $x$-direction.}\label{fig:squaretube}
\end{figure}

The defect and surplus angles of the grid lines can be oriented along the direction of propagation of the grid, i.e. the positive $u$ direction. The cuts of the defect/surplus angles then divide the $u>0$ side of the grid in square ``tubes'' of spacetime (see figure \ref{fig:squaretube}). The Minkowski metric from $u<0$ side of the grid can be continued into each tube. The constant $u$ and $v$ slices in each tube start out as a square at the base ($u=0$) of the tube. As $u$ increases the square is elongated in the $x$ direction and compressed in the $y$ direction.

We now construct coordinates on each tube, which can be patched together to one coordinate system on the $u>0$ side of the spacetime. The space
\begin{equation}
T_{m,n}=\RR_{>0}\times\RR\times [\epsilon(m-\tfrac{1}{2}),\epsilon(m+\tfrac{1}{2})]\times [\epsilon(n-\tfrac{1}{2}),\epsilon(n+\tfrac{1}{2})]
\end{equation}
can be mapped to the tube behind the face of the grid centered on $x=\epsilon m$ and $y=\epsilon n$ by the mapping,
\begin{equation}
\begin{aligned}
 u &\mapsto u,\\
 v &\mapsto v,\\
 x &\mapsto \epsilon m+(1+\frac{\alpha}{\epsilon} u)\bhh{x-\epsilon m},\\
 y &\mapsto \epsilon n+(1-\frac{\alpha}{\epsilon} u)\bhh{y-\epsilon n}.
\end{aligned}
\end{equation}
The metric induced on $T_{m,n}$ by this mapping is,
\begin{equation}\label{eq:Tmnmetric}
\md s^2 = 2\md u \md v+
\hh{(1+\tfrac{\alpha}{\epsilon} u)\md x + \tfrac{\alpha}{\epsilon}(x-\epsilon m)\md u}^2+
\hh{(1-\tfrac{\alpha}{\epsilon} u)\md y -\tfrac{\alpha}{\epsilon}(y-\epsilon n)\md u}^2.
\end{equation}
Together, the patches $T_{m,n}$ cover the entire $u>0$ side of the grid. The metric \eqref{eq:Tmnmetric} may be viewed as a metric on the union of all $T_{m,n}$, $\RR_{>0}\times\RR^3$, if $m$ and $n$ are interpreted as functions of $x$ and $y$ respectively. That is,\footnote{Here $\floor{x}$ denotes the floor function, which returns the largest integer smaller than or equal to $x$.}
\begin{equation}
\begin{aligned}
m(x) &= \floor{x/\epsilon + 1/2},\\
n(y) &= \floor{y/\epsilon + 1/2}.
\end{aligned}
\end{equation}
These functions are not continuous, so we need to worry whether \eqref{eq:Tmnmetric} describes a well-defined geometry on the whole $u>0$ half-space. The appropriate condition \cite{Clarke:1987,Mars:1993mj} is that the metric induced on the common boundary of two patches is the same in both patches.

On the $x=\epsilon(m+\tfrac{1}{2})$ boundary the metric on $T_{m,n}$ induces the 3-dimensional metric,
\begin{equation}
\md s^2 = 2\md u \md v+
\hh{(1-\tfrac{\alpha}{\epsilon} u)\md y -\tfrac{\alpha}{\epsilon}(y-\epsilon n)\md u}^2.
\end{equation}
On the other side of the boundary the metric on the $T_{m+1,n}$, can easily be seen to induce the same 3-dimensional metric. Similarly, we find that the 3-dimensional metrics induced on the $y=\epsilon(n+\tfrac{1}{2})$ boundaries match, and hence that the metric \eqref{eq:Tmnmetric} is well-defined on the whole $u>0$ half-space.

Moreover, the 3-dimensional metric induced on the $u=0$ hyperplane is
\begin{equation}
\md s^2 = \md x^2+  \md y^2,
\end{equation}
which agrees with the metric induced from the Minkowski side. Consequently, the metric \eqref{eq:lcMinmetric} for $u<0$ and the metric \eqref{eq:Tmnmetric} for $u>0$ together describe a well-defined geometry on the whole spacetime, given by the metric
\begin{multline}
\md s^2 = 2\md u \md v+
\hh{(1+\tfrac{\alpha}{\epsilon} u\theta(u))\md x + \tfrac{\alpha}{\epsilon}(x-\epsilon m)\theta(u)\md u}^2\\+
\hh{(1-\tfrac{\alpha}{\epsilon} u\theta(u))\md y -\tfrac{\alpha}{\epsilon}(y-\epsilon n)\theta(u)\md u}^2,
\end{multline}
where $\theta(u)$ is the Heaviside stepfunction.

This metric should describe the geometry of our piecewise flat configuration of defects. As a check, we can calculate the Riemann tensor of this metric. Since the metric is discontinuous along the boundaries of the patches, we need to treat the curvature tensors as tensor distributions as described in \cite{Mars:1993mj}. The upshot of this is that in a case like this we can and should treat the metric components as distributions rather than normal functions. Using the fact that in a distributional sense
\begin{align}
\frac{\md}{\md x} \floor{x} &= \sum_{i\in\ZZ} \delta(x - i)\equiv \Delta(x),\quad\text{and}\\
\frac{\md^2}{\md u^2} u\theta(u) &= \delta(u),
\end{align}
we find that the Riemann tensor is given by\footnote{For notational convenience we denote the tensor product of a 2-form with itself by a square, i.e. $(\md u\wedge\md y)^2 = (\md u\wedge\md y)\otimes(\md u\wedge\md y)$.}
\begin{equation}
R = 4\alpha \delta(u)\bhh{\Delta_\epsilon(y+\tfrac{\epsilon}{2})(\md u\wedge\md y)^2 -\Delta_\epsilon(x+\tfrac{\epsilon}{2})(\md u\wedge\md x)^2},
\end{equation}
where $\Delta_\epsilon(x)$ is the Dirac comb with period $\epsilon$, i.e.
\begin{equation}
\Delta_\epsilon(x)\equiv  \sum_{i\in\ZZ} \delta(x - \epsilon i).
\end{equation}
We observe that the Riemann tensor indeed vanishes everywhere, except on the grid of defects where it has singular delta peaks as expected. We have therefore found a metric for our piecewise flat spacetime consisting of a grid of lightlike defects.

Our next step is to find the continuum limit of this spacetime. For this we will send the grid size $\epsilon$ to zero while keeping the density of defects (approximately) constant. The number of defects in a unit area of the grid scales inversely proportionately to $\epsilon$, hence the amplitude of the defects $\alpha$ should scale linearly with $\epsilon$ to keep the total amplitude per unit area constant. Inserting $\alpha=\epsilon \alpha_0$ in the above metric and taking the limit of $\epsilon$ goes to zero, we find
\begin{equation}\label{eq:impulsivewave}
\md s^2 = 2\md u \md v+
\hh{1+\alpha_0 u\theta(u)}^2 \md x^2 + \hh{1-\alpha_0 u\theta(u)}^2 \md y^2,
\end{equation}
where $\theta(u)$ is the Heaviside step function. The Riemann tensor (distribution) is given by
\begin{equation}
R = 4\alpha_0 \delta(u)\bhh{(\md u\wedge\md y)^2 -(\md u\wedge\md x)^2},
\end{equation}
from which the Einstein tensor may be obtained, which is seen to vanish. Consequently, the metric \eqref{eq:impulsivewave} is an exact solution of the vacuum Einstein equation, a gravitational wave consisting of a single plane wavefront. Waves of this type are called impulsive  gravitational plane waves, and were first considered by Penrose \cite{Penrose:1972}.

\section{General gravitational plane waves}\label{sec:genplanewaves}
We have succeeded in explicitly constructing a family of exact defect configurations that converges to a specific exact gravitational wave solution of general relativity in the limit that the configuration becomes continuous. In the construction we made some explicit choices for the sake of simplicity. We chose to align the grid with the coordinate axes and as a result described a plane wavefront with ``$+$'' polarization. Rotating the grid simply results in a rotation of the $\md x$ and $\md y$ terms in \eqref{eq:impulsivewave}, producing a wavefront with a different polarization. For example, rotating the grid by $\pi/4$ produces a wavefront with ``$\times$'' polarization. The result always is an impulsive plane wave with constant polarization. 

We could also take a different geometry for the grid. Equation \eqref{eq:gridjunction} shows that the components of the grid do not need to be perpendicular, nor do the components have to have matching amplitudes. As long as the curvature density of the positive curvature defects is equal to the curvature density of the negative curvature defects, the Einstein tensor will vanish in the continuum limit. Different geometries will not produce any wavefronts that could not be produced by a square grid.

Gravitational waves with longer wave packets may be constructed as a succession of wavefronts. Each wavefront contributes to the Riemann  curvature in the following way,
\begin{equation}
R = \alpha \delta(u-u_0)\bb{\bhh{\md u\wedge(\cos\phi\;\md x+\sin\phi\;\md y)}^2 -\bhh{\md u\wedge(\sin\phi\;\md x-\cos\phi\;\md y)}^2},
\end{equation}
where $\phi$ is the direction of the polarization of the wavefront.

A continuous succession of wavefronts with $\alpha$ and $\phi$ varying with $u$ therefore produces a wave with curvature,
\begin{equation}
R = \alpha(u)\bb{\bhh{\md u\wedge(\cos\phi(u)\;\md x+\sin\phi(u)\;\md y)}^2 -\bhh{\md u\wedge(\sin\phi(u)\;\md x -\cos\phi(u)\;\md y)}^2}.
\end{equation}
This is the most general curvature produced by exact gravitational plane waves of the type described by Bondi et al. in \cite{bondi:1959}.

\section{Other gravitational waves}\label{sec:otherwaves}
We have shown that it is possible to construct general gravitational plane waves in our piecewise flat model of gravity. Since a general gravitational wave will locally resemble a plane wave, this suggests that it should be possible to construct any gravitational wave. This is further supported by the observations that any gravitational wave may be obtained in the linear limit of the model \cite{meent:2010b}, and that its is possible to find a piecewise flat approximation to any solution of general relativity \cite{CMS:1984}. Although this is all very suggestive, it is not enough to conclude that any gravitational wave may be constructed in our piecewise flat model of gravity. It is therefore instructive to explicitly examine the construction of a non-planar wave in our model and see what new issues arise.

As an example, we will consider cylindrical gravitational waves. In the case of cylindrical symmetry the solutions of the vacuum Einstein equation can found by solving a single linear differential equation\cite{Einstein:1937qu,Weber:1957}. Consequently, the complete set of possible solutions is known. In particular, it is known that it is impossible to form isolated impulsive wavefronts. Instead, impulsive wavefronts leave a wake of backscattered gravitational waves. Explicit solutions with impulsive wavefronts \cite{Alekseev:1991kz} show that the form of this backscattered component can be quite complicated. This suggests that it may be too much to hope for a simple description of the backscattered component in terms of a piecewise flat geometry.\footnote{This says nothing about whether a piecewise flat description exists. We simply observe that constructing one will be very complicated.} Instead we will focus on constructing an impulsive cylindrical gravitational wave front, while we allow the backscattered component to be any defect configuration, i.e. not necessarily one with vanishing energy--momentum.

\begin{figure}[tbp]
\centering
\includegraphics[width=120mm]{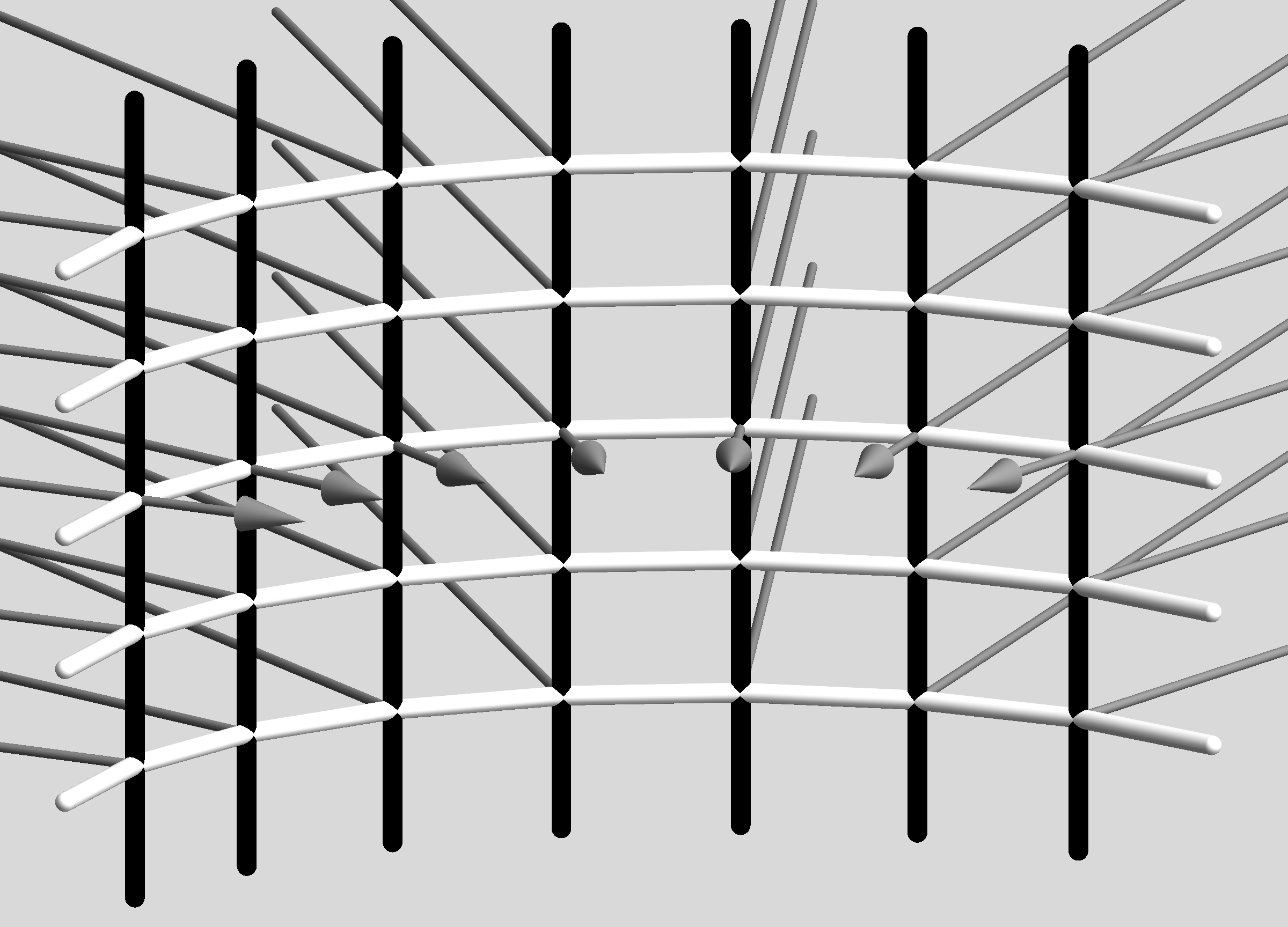}
\caption{To first order, a piecewise flat cylindrical wave front should consist of axial (black) and azimuthal (white) defects moving inwards in the radial direction. The junction conditions imply that any additional radial defect (gray) should be present at each junction.}\label{fig:cylindricalfront}
\end{figure}

Let us consider a single ingoing cylindrical impulsive wavefront located at $u=0$,\footnote{We use cylindrical lightcone coordinates defined by  $u=\tfrac{r+t}{\sqrt{2}}$ and $v=\tfrac{r-t}{\sqrt{2}}$.} which propagates into a Minkowski background (i.e. the Riemann tensor vanishes for $u<0$). In the limit of large radii ($v\rightarrow \infty$), the wavefront will resemble an impulsive plane wave. Our first order guess for the piecewise flat cylindrical wavefront therefore consists of a grid of perpendicular massless defects with positive and negative energy oriented in the axial and azimuthal directions and moving in the radial direction (see figure \ref{fig:cylindricalfront}). 

Curving a flat grid to a cylinder introduces two new complications. The first
is that due to simple geometric considerations the velocity of the azimuthal defects must be smaller than the velocity of the axial defects. The azimuthal defects, therefore, are not quite massless. However, as the grid is made finer this effect disappears and the azimuthal defects again become massless in the continuum limit. The second effect is that the four defects of the grid meeting at each junction do not lie in one plane. A consequence of this is that the junction condition \eqref{eq:ewjc} cannot be satisfied by these four defects. To satisfy \eqref{eq:ewjc} a fifth defect pointing in the outward radial direction must be added to each junction (depicted as the gray lines in figure \ref{fig:cylindricalfront}). Just like general relativity, the piecewise flat model does not allow isolated cylindrical wavefronts. 

Repeating the steps of section \ref{sec:planewavefronts} we can find the piecewise flat metric associated to this configuration and calculate the continuum limit. The result is
\begin{equation}
\md s^2 = 2\md u \md v + \frac{(u+v+\alpha u\theta(u))^2}{2}\md\phi^2 + (1-\alpha u \theta(u))^2 \md z^2.
\end{equation}
The corresponding energy--momentum tensor is
\begin{equation}
T=\alpha \frac{v-1}{v} \delta(u) \md u^2 - 2\alpha \frac{1}{(1-\alpha u)(v+(1+\alpha)u)}\theta(u)\md u\md v.
\end{equation}
As expected the energy--momentum vanishes in the $u<0$ region, but does not in the `backscattered' $u>0$ region due to the presence of the radial defects. However, the energy--momentum also has a non-vanishing delta peak at $u=0$ for $v\neq 1$, i.e. it fails to model a cylindrical impulsive gravitational wavefront. This failure is the result of the two singular components of the Riemann tensor,
\begin{equation}
R = 4 \alpha \delta(u) (\md u\wedge\md z)^2  -4\frac{\alpha}{v}\delta(u)(\md u\wedge\frac{v}{\sqrt{2}}\md\phi)^2 +\theta(u)\bhh{\cdots},
\end{equation}
scaling differently with $v$. The expected behaviour --- as can be deduced from linear cylindrical waves, or by expanding the exact cylindrical solution in \cite{Alekseev:1991kz} for small values of $u$ --- is for both curvature components to scale inversely with $\sqrt{v}$.

On the level of the piecewise flat geometry, this happens because the number of axial defects stays constant while the radius changes, causing their density to scale inversely with the radius. The density of the azimuthal defects, on the other hand, stays constant. To change the scaling of the impulsive front, the energy densities of the defects must change as the defects move inward. The only way this can happen is if the defects periodically emit new defects into the backscattered region.

\begin{figure}[tbp]
\centering
\includegraphics[width=120mm]{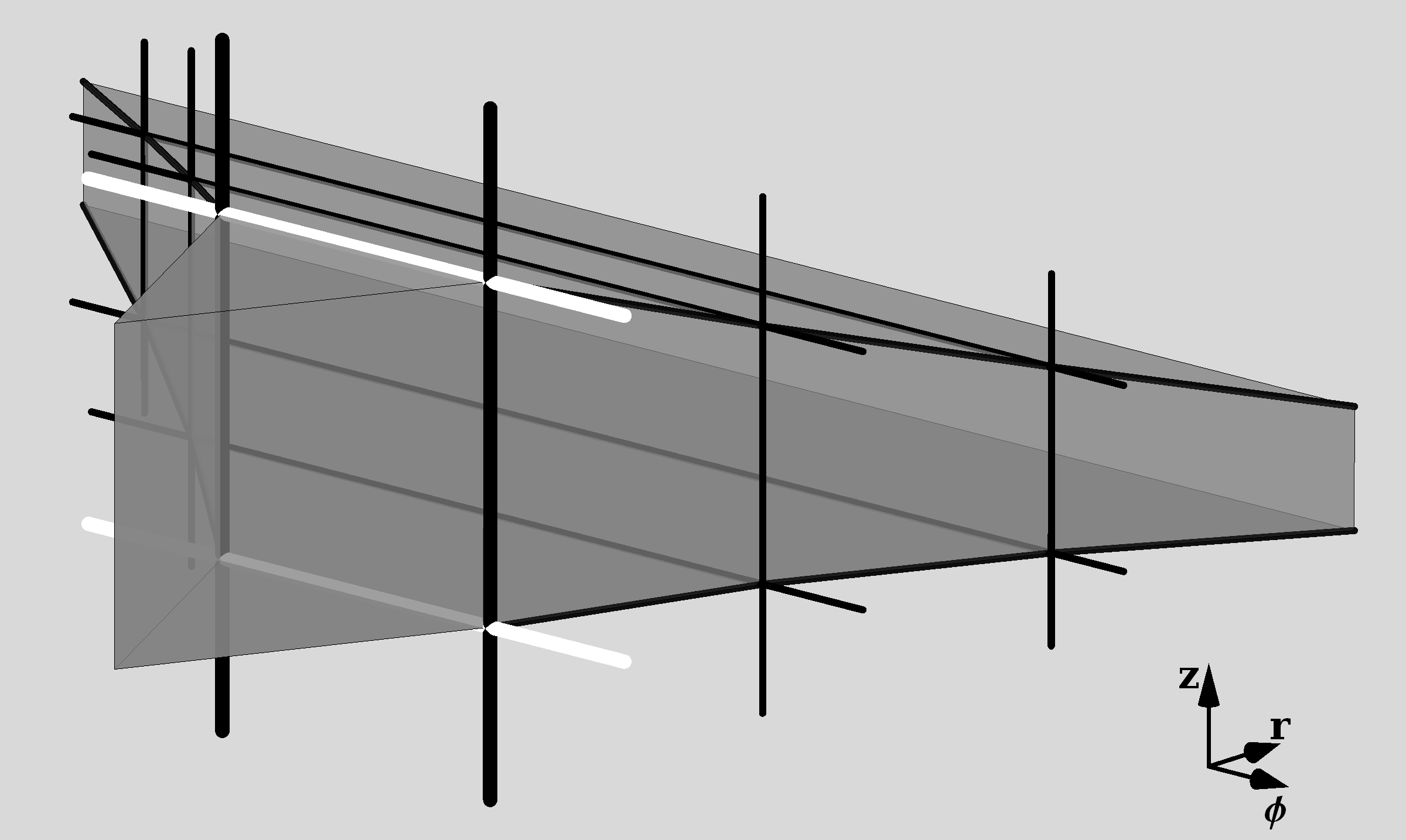}
\caption{The geometry of a single cell of the piecewise flat geometry approximating an impulsive cylindrical gravitational wave front.}\label{fig:cylindricaltube}
\end{figure}

We implement this by having the ingoing impulsive front emit an outward moving lightlike front of perpendicular defects. A single cell of the piecewise flat geometry then takes shape shown in figure \ref{fig:cylindricaltube}. By adjusting the frequency and the energies of the emitted perpendicular components we can achieve any scaling with $v$ that we want. To achieve the proper scaling for the curvature of a cylindrical wave front, we choose the emitted components such that the metric in the continuum limit becomes,
\begin{equation}
\md s^2 = 2\md u \md v + \frac{(u+v+\alpha\sqrt{v} u\theta(u))^2}{2}\md\phi^2 + (1-\alpha \frac{u}{\sqrt{v}} \theta(u))^2 \md z^2.
\end{equation}
The singular part of the corresponding energy--momentum tensor is
\begin{equation}
T = 0 +\theta(u)\bhh{\cdots}.
\end{equation}
WThe impulsive wavefront has thus vanished from the energy--momentum tensor. Yet, the singular part of the Riemann curvature,
\begin{equation}
R = 4 \frac{\alpha}{\sqrt{v}} \delta(u) (\md u\wedge\md z)^2  -4\frac{\alpha}{\sqrt{v}}\delta(u)(\md u\wedge\frac{v}{\sqrt{2}}\md\phi)^2 +\theta(u)\bhh{\cdots},
\end{equation}
does not vanish, and has the expected scaling for a cylindrical impulsive wavefront. The described configuration indeed contains a cylindrical impulsive wavefront.

The described piecewise flat geometry is just one way an impulsive cylindrical gravitational wavefront could evolve in our piecewise flat model of gravity. Recall that the model fundamentally describes gravity coupled to matter. As discussed in our previous articles \cite{hooft:2008, meent:2010}, the dynamics of the model is incomplete.\footnote{This is related to the different ways in which a piecewise flat geometry can be continued after the collision of two defects.} It needs to be complemented by additional rules which relate hoe the matter content interacts, much like general relativity needs to be complemented by equations of motion for the matter content. Different dynamical completions of the model lead to different evolutions of cylindrical wavefronts. The specific evolution described here corresponds a matter interaction where interacting gravitational waves can produce other types of matter.  

One might ask if there exists a dynamical completion of the model that corresponds to pure gravity with no matter present at all. Of course, this is of academic interest only since the physical world also contains matter. In such a completion the backscattered component of the evolving cylindrical wavefront must also satisfy the vacuum Einstein equation. This requires the addition of additional defects to the configuration. These will break the remaining residual symmetry of the piecewise flat geometry. As a consequence it will no longer be possible to describe the geometry as an infinite collection of identical piecewise flat cells, substantially adding to the complexity of the description. This breaking of symmetry indicates that cylindrical gravitational waves may not be the best setting to further study the dynamical completion of the piecewise flat model.

Nonetheless, the study of the interaction of gravitational waves in the piecewise flat model may provide essential guiding insights for dynamically completing the model. A more promising setting for this study is the interaction of two plane wavefronts. This problem is well documented in the case of general relativity coupled to various types of matter \cite{Khan:1971vh,Griffiths:1991,Barrabes:2002yg,Stephani:2003}. Moreover, the intrinsic plane symmetry is much more suited to a piecewise flat geometry. Furthermore, it is well known that the future of plane wave collisions can contain regions which are locally diffeomorphic to black hole solutions \cite{Khan:1971vh, Stephani:2003}. Studying plane wave collisions may therefore offer crucial insight in one of the major open problems of the piecewise flat model; the construction of non-trivial stationary vacuum solutions such as black holes. However, the study of dynamical completions of the model is a major project on its own and will not be attempted here.

\section{Conclusions}
In this article we have extended our previous linear result \cite{meent:2010b} for obtaining gravitational waves in our piecewise flat model for gravity in 3+1 dimensions to an explicit construction of a family of exact piecewise flat configurations that approaches exact plane wave solutions of general relativity. This suggests that the construction of general gravitational waves may be possible. As an example, we have constructed the evolution of a cylindrical gravitational wavefront allowing the backscattering to produce other types of matter. 

Understanding the construction of gravitational waves is an important step towards showing that the proposed model can indeed serve as description of gravity in 3+1 dimensions. Further study of the interactions of gravitational waves may provide essential guiding insights for completing the incomplete dynamics of the model.

\section*{Acknowledgments}
The author would like to thank his advisor, Gerard 't Hooft, for many fruitful discussions and useful comments.
\bibliographystyle{../utcaps}
\bibliography{../thesis/bib/thesis}

\providecommand{\href}[2]{#2}\begingroup\raggedright\begin{thebibliography}{10}

\bibitem{meent:2010b}
M.~van~de Meent, ``{Piecewise flat gravitational waves},''
  \href{http://dx.doi.org/10.1088/0264-9381/28/7/075005}{{\em Classical and
  Quantum Gravity} {\bf 28} (2011)  075005},
  \href{http://arxiv.org/abs/1012.1991}{{\tt arXiv:1012.1991 [gr-qc]}}.

\bibitem{hooft:2008}
G.~'t~Hooft, ``{A locally finite model for gravity},''
  \href{http://dx.doi.org/10.1007/s10701-008-9231-3}{{\em Foundations of
  Physics} {\bf 38} (2008)  733--757},
  \href{http://arxiv.org/abs/0804.0328}{{\tt arXiv:0804.0328 [gr-qc]}}.

\bibitem{meent:2010}
M.~van~de Meent, ``{Collisions in piecewise flat gravity in 3+1 dimensions},''
  \href{http://dx.doi.org/10.1088/0264-9381/27/14/145003}{{\em Classical and
  Quantum Gravity} {\bf 27} (2010)  145003},
  \href{http://arxiv.org/abs/1002.3708}{{\tt arXiv:1002.3708 [gr-qc]}}.

\bibitem{Witten:1988hc}
E.~Witten, ``2+1 dimensional gravity as an exactly soluble system,''
  \href{http://dx.doi.org/10.1016/0550-3213(88)90143-5}{{\em Nuclear Physics B}
  {\bf 311} (1988)  46--78}.

\bibitem{Deser:1989}
S.~Deser, J.~McCarthy, and Z.~Yang, ``Polynomial formulations and
  renormalizability in quantum gravity,''
  \href{http://dx.doi.org/10.1016/0370-2693(89)90723-5}{{\em Physics Letters B}
  {\bf 222} (1989)  61--65}.

\bibitem{Deser:1983tn}
S.~Deser, R.~Jackiw, and G.~'t~Hooft, ``{Three-dimensional Einstein gravity:
  Dynamics of flat space},''
  \href{http://dx.doi.org/10.1016/0003-4916(84)90085-X}{{\em Annals of Physics}
  {\bf 152} (1984)  220--235}.

\bibitem{Waelbroeck:1994iy}
H.~Waelbroeck, ``{Canonical quantization of (2+1)-dimensional gravity},''
  \href{http://dx.doi.org/10.1103/PhysRevD.50.4982}{{\em Physics Review D} {\bf
  50} (1994)  4982--4992}, \href{http://arxiv.org/abs/gr-qc/9401022}{{\tt
  arXiv:gr-qc/9401022}}.

\bibitem{hooft:1996uc}
G.~'t~Hooft, ``{Quantization of point particles in 2+1 dimensional gravity and
  space-time discreteness},''
  \href{http://dx.doi.org/10.1088/0264-9381/13/5/018}{{\em Classical and
  Quantum Gravity} {\bf 13} (1996)  1023--1040},
  \href{http://arxiv.org/abs/gr-qc/9601014}{{\tt arXiv:gr-qc/9601014}}.

\bibitem{regge:1961}
T.~Regge, ``General Relativity without Coordinates,''
  \href{http://dx.doi.org/10.1007/BF02733251}{{\em Nuovo Cimento A} {\bf 19}
  (1961)  558--571}.

\bibitem{RW:2000}
T.~Regge and R.~M. Williams, ``{Discrete structures in gravity},''
  \href{http://dx.doi.org/10.1063/1.533333}{{\em Journal of Mathematical
  Physics} {\bf 41} (2000)  3964--3984},
  \href{http://arxiv.org/abs/gr-qc/0012035}{{\tt arXiv:gr-qc/0012035}}.

\bibitem{Weber:1957}
J.~Weber and J.~A. Wheeler, ``{Reality of the cylindrical gravitational waves
  of Einstein and Rosen},''
  \href{http://dx.doi.org/10.1103/RevModPhys.29.509}{{\em Reviews of Modern
  Physics} {\bf 29} (1957)  509--515}.

\bibitem{bondi:1959}
H.~Bondi, F.~A.~E. Pirani, and I.~Robinson, ``{Gravitational waves in general
  relativity. III. Exact plane waves},'' {\em Proceedings of the Royal Society
  A} {\bf 251} (1959)  519--533, \href{http://arxiv.org/abs/100727}{{\tt
  100727}}.

\bibitem{Clarke:1987}
C.~Clarke and T.~Dray, ``Junction conditions for null hypersurfaces,''
  \href{http://dx.doi.org/10.1088/0264-9381/4/2/010}{{\em Classical and Quantum
  Gravity} {\bf 4} (1987)  265}.

\bibitem{Mars:1993mj}
M.~Mars and J.~M. Senovilla, ``{Geometry of general hypersurfaces in
  space-time: Junction conditions},''
  \href{http://dx.doi.org/10.1088/0264-9381/10/9/026}{{\em Classical and
  Quantum Gravity} {\bf 10} (1993)  1865--1897},
  \href{http://arxiv.org/abs/gr-qc/0201054}{{\tt arXiv:gr-qc/0201054}}.

\bibitem{Penrose:1972}
R.~Penrose, ``The geometry of impulsive gravitational waves,'' in {\em General
  Relativity, Papers in Honour of JL Synge}, pp.~101--15.
\newblock Clarendon Press, 1972.

\bibitem{CMS:1984}
J.~Cheeger, W.~M{\"u}ller, and R.~Schrader, ``{On the curvature of piecewise
  flat spaces},'' \href{http://dx.doi.org/10.1007/BF01210729}{{\em
  Communications in Mathematical Physics} {\bf 92} (1984)  405--454}.

\bibitem{Einstein:1937qu}
A.~Einstein and N.~Rosen, ``{On gravitational waves},''
  \href{http://dx.doi.org/10.1016/S0016-0032(37)90583-0}{{\em Journal of the
  Franklin Institute} {\bf 223} (1937)  43--54}.

\bibitem{Alekseev:1991kz}
G.~A. Alekseev and J.~Griffiths, ``{Exact solutions for gravitational waves
  with cylindrical, spherical and toroidal wavefronts},''
  \href{http://dx.doi.org/10.1088/0264-9381/13/8/014}{{\em Classical and
  Quantum Gravity} {\bf 13} (1996)  2191--2209}.

\bibitem{Khan:1971vh}
K.~Khan and R.~Penrose, ``{Scattering of two impulsive gravitational plane
  waves},'' \href{http://dx.doi.org/10.1038/229185a0}{{\em Nature} {\bf 229}
  (1971)  185--186}.

\bibitem{Griffiths:1991}
J.~Griffiths, {\em {Colliding plane waves in general relativity}}.
\newblock Oxford University Press, 1991.
\newblock \url{http://www-staff.lboro.ac.uk/~majbg/jbg/book.html}.

\bibitem{Barrabes:2002yg}
C.~Barrabes and P.~Hogan, ``{A class of collisions of plane impulsive light
  like signals in general relativity},''
  \href{http://dx.doi.org/10.1142/S0218271802002116}{{\em International Journal
  of Modern Physics D} {\bf 11} (2002)  933--945},
  \href{http://arxiv.org/abs/gr-qc/0209112}{{\tt arXiv:gr-qc/0209112}}.

\bibitem{Stephani:2003}
H.~Stephani, D.~Kramer, M.~MacCallum, C.~Hoenselaers, and E.~Herlt, {\em Exact
  Solutions of {Einstein's} Field Equations}.
\newblock Cambridge University Press, second~ed., 2003.

\end{thebibliography}\endgroup
\end{document}